\documentclass[11pt]{article}
\usepackage{jheppub}
\usepackage{amsmath}
\usepackage{amssymb}
\usepackage{braket}
\usepackage{bbold}
\usepackage{pifont}
\usepackage{color}

\newcommand{\beqn}{\begin{eqnarray}}
\newcommand{\eeqn}{\end{eqnarray}}
\newcommand{\dd}{\mathrm{d}}
\newcommand{\nn}{\nonumber}

\newcommand{\fmn}{f_{\mu\nu}}

\newcommand{\te}{{\tilde{e}}}
\newcommand{\be}{{\bar{e}}}			
\newcommand{\tE}{{\varphi}}	
\newcommand{\EE}{\mathcal{E}}	
\newcommand{\gmn}{g_{\mu\nu}}
\newcommand{\bmn}{B_{\mu\nu}}

\title{Gravity with antisymmetric components}
\author{Chrysoula Markou, Felix J. Rudolph \& Angnis~Schmidt-May}
\affiliation{Max-Planck-Institut f\"ur Physik (Werner-Heisenberg-Institut)\\
F\"ohringer Ring 6, 80805 Munich, Germany}
\emailAdd{cmarkou@mpp.mpg.de, frudolph@mpp.mpg.de, angnissm@mpp.mpg.de}

\abstract{This work proposes a new gravitational theory formulated in terms of the vierbein field.
The vierbein contains components which can be shifted by local Lorentz transformations and therefore
do not show up in the spacetime metric. These components are given dynamics and become physical in 
our setup. They enter the massless theory in the form of an antisymmetric tensor field which
makes the action reminiscent of the bosonic sector of supergravity. We then demonstrate
that both the metric and the antisymmetric tensor can be made massive by adding a potential term for the
vierbein. The form of this mass potential is inspired by ghost-free massive gravity. We confirm the
absence of additional and potentially pathological degrees of freedom in an ADM analysis.
However, at the linearized level around maximally symmetric solutions, the fluctuation of the 
antisymmetric tensor has a tachyonic mass pole.}

\begin{document} 

\begin{flushright}
\hfill{MPP-2018-288} \vspace{20mm}
\end{flushright}
\maketitle
\flushbottom

\section{Introduction}

\subsection{Motivation}

General relativity is the standard theory for gravitational interactions.
It possesses a mathematically well-defined interpretation in terms of Riemannian
geometry and describes the dynamics of a metric $\gmn$ on a Riemannian manifold~$\mathcal{M}$.
The theory is well-tested over a large range of different energies and, from an experimental or
observational point of view, there may be little motivation to modify or extend its 
unique and exceptionally beautiful structure. On the other hand, general relativity
is not a fundamental quantum field theory because it is not renormalizable. 
It can be treated as an effective field theory valid at low energies which requires
an ultraviolet completion at the Planck scale the latest. A well-studied and 
promising candidate for the quantum theory of gravity at high energies is string theory,
in which the notion of a point particle is replaced by a one-dimensional string.

The spacetime metric $\gmn$ arises as part of the massless bosonic excitations of the closed string, 
which in addition contain the scalar dilaton $\phi$ and the antisymmetric tensor $\bmn$, 
also known as the Kalb-Ramond field~\cite{Kalb:1974yc}.
At low energies, string theory reduces to supergravity in 10 spacetime dimensions whose 
bosonic field content is precisely given by the massless fields $\gmn$, $\bmn$ and~$\phi$.
For many purposes, one can consistently ignore the effects of the dilaton and work in an 
approximation where it is constant.\footnote{For example, one can assume that the value of the dilaton 
has been stabilized by means of a flux compactification (for a comprehensive review of the latter see \cite{Grana:2005jc}).} 
We therefore restrict our discussion in the following to the tensor fields.

The geometry of quantum gravity is not known, but string theory can provide us with hints on its nature. 
For instance, strings are subject to duality symmetries which do not have an obvious interpretation
in standard field theory but are crucial for quantum consistency.
Some aspects of these dualities can be captured using the mathematical 
concept of \textit{generalized geometry}~\cite{Gualtieri:2003dx, Hitchin:2004ut}.
Attempts to incorporate the T-duality symmetry of closed strings into 10-dimensional supergravity 
inspired the creation of \textit{double field theory (DFT)}~\cite{Siegel:1993xq,Siegel:1993th, Hull:2009mi} 
(see also the reviews~\cite{Aldazabal:2013sca, Berman:2013eva, Hohm:2013bwa}).
DFT is formulated on a doubled space obtained by augmenting the coordinates of regular 
supergravity with additional spatial coordinates associated with the winding modes of the string. 
The metric $\gmn$ and Kalb-Ramond field $\bmn$ are combined into a single symmetric tensor 
corresponding to a ``generalized metric'' of the doubled space. 
Thus $\gmn$ and $\bmn$ are treated on an equal footing and are both seen as 
part of the geometry. For an extensive analysis of the mathematical structure of the 
doubled geometry and its relation to generalized geometry see 
for example \cite{Freidel:2017yuv, Freidel:2018tkj}.

String theory also contains other extended objects such as 
D-branes whose dynamics can be described by the 
Dirac-Born-Infeld action~\cite{Leigh:1989jq, Polchinski:1995mt}. 
This action is another example of containing the massless fields 
in the form of a single tensor, namely the linear combination $\gmn+\bmn$.

In this work we propose a new way to combine the degrees of freedom of the fields $\gmn$ and $\bmn$
into a single object with a well-known geometrical interpretation.  This object is taken to be
 the vierbein field $e^a_{~\mu}$ defined via $\gmn = e^a_{~\mu}\eta_{ab}e^b_{~\nu}$.
Since the metric is invariant under local Lorentz transformations of the vierbein,
$e^a_{~\mu}\mapsto\Lambda^a_{~b}e^b_{~\mu}$ with $\Lambda^a_{~b}\eta_{ac}\Lambda^c_{~d}=\eta_{bd}$,
$\gmn$ depends only on 10 of the 16 components in $e^a_{~\mu}$.\footnote{For concreteness, we work
in $D=4$ spacetime dimensions throughout this paper, but our results straightforwardly generalize to
arbitrary $D$, where $e^a_{~\mu}$ contains $D^2$ and $\gmn$ contains $D(D+1)/2$ components.} 
In general relativity, the remaining
6 vierbein components are unphysical. Our idea is to make them dynamical and identify them with the 6
components of the antisymmetric tensor $\bmn$. This nontrivial way of re-packaging the field components 
into a single matrix is not directly inspired by string theory but rather follows a bottom-up approach, 
starting from the simple setup of general relativity coupled to an antisymmetric tensor $\bmn$. 

In fact, there exists already a well-studied set of theories for which the additional Lorentz components 
in the vierbein are non-dynamical but not unphysical, at least not initially. 
These theories are the vierbein formulations of \textit{nonlinear massive gravity}~\cite{deRham:2010kj}, 
\textit{bimetric gravity}~\cite{Hassan:2011zd} 
and \textit{multimetric gravity}~\cite{Hinterbichler:2012cn}
(for reviews see~\cite{Hinterbichler:2011tt, deRham:2014zqa, Schmidt-May:2015vnx}).
In these setups, the 6 extra vierbein components are determined by solving a set of constraint equations.
Generically, these constraints eventually require the additional degrees of freedom to vanish.
This feature is crucial for the absence of ghost instabilities in these theories~\cite{deRham:2015cha}
and therefore the extra components cannot be made to remain physical in a straightforward way. The idea 
to make the additional Lorentz components dynamical in a way that avoids the ghost was already mentioned 
in Ref.~\cite{Gabadadze:2013ria} where, however, the suggested parametrization was slightly different from our setup. 

A discussion of a dynamical antisymmetric part of the vierbein fluctuation in the context of linearized teleparallel theories 
can be found in section 4.6 of the encyclopaedic book by Ort\'in \cite{Ortin:2004ms}. 
It is also worth mentioning that a previous attempt to extend general relativity 
by additional antisymmetric components was made by introducing
Hermitian metrics~\cite{Moffat:1978tr, Chamseddine:2000zu, Chamseddine:2012gh, Chamseddine:2010rv}.
For more work on gravity in the presence of an independent $\bmn$ 
field, see e.g.~Ref.~\cite{Altschul:2009ae, Lau:2013tda,Aashish:2018jzo,Aashish:2018aqn,Aashish:2018lhv}.

\subsection*{Summary of results}

We extract the additional components of the vierbein that do not appear in the metric
$\gmn= e^a_{~\mu}\eta_{ab}e^b_{~\nu}$ by introducing a second 
(non-dynamical) vierbein $\te^a_{~\mu}$ and defining
$\bmn$ as the antisymmetric part of the matrix $e^a_{~\mu}\eta_{ab}\te^b_{~\nu}$.
These definitions are inserted into the action for the massless fields, which contains an Einstein-Hilbert term 
for the metric and a standard quadratic kinetic term for the antisymmetric tensor. 
We demonstrate that the formulation of the massless theory in terms of the vierbein field is entirely equivalent 
to the tensor formulation with independent $\gmn$ and $\bmn$. 

We then extend the massless action by a set of vierbein interactions
whose structure is motivated by massive gravity. As a highly nontrivial result,
we find that these interactions make both fields $\gmn$ and $\bmn$ massive 
without introducing any further degrees of freedom into the theory (which
usually are pathological). This is verified both at the linear level 
where the fields decouple and in a fully nonlinear ADM analysis.
The antisymmetric fluctuation around maximally symmetric backgrounds has a tachyonic
mass pole. This implies that the vacuum solution with $\bmn=0$ (which is also the massive
gravity vacuum) is unstable.

The resulting vierbein theory thus yields an entirely new set of ghost-free gravitational
interactions, which can both be seen as an extension of massive gravity and as an extension of
the massless action for the supergravity fields $\gmn$ and $\bmn$. It remains an open question
whether the tachyonic behavior of the antisymmetric fluctuation around the massive gravity solution
can be cured.

\subsection*{Conventions}
Throughout this paper we will use the following conventions. We are working in 4 dimensions and the signature of the spacetime metric is $(-,+,+,+)$. The spacetime indices are denoted by Greek letters $\mu,\nu$ whereas for the Lorentz indices we use Latin letters $a,b$. These two types of indices are related via the vierbein $e^a_{~\mu}$ and its inverse, $e^{\mu}_{~a}$. Indices are raised and lowered by $\gmn$ and $\eta_{ab}$ and their inverses respectively. Brackets to denote symmetrization and antisymmetrization over indices are defined as $T_{\mu\nu} = T_{(\mu\nu)} + T_{[\mu\nu]}$ with $T_{(\mu\nu)} = \frac12(T_{\mu\nu} + T_{\nu\mu})$ and $T_{[\mu\nu]} = \frac12(T_{\mu\nu} - T_{\nu\mu})$.

\subsection{The standard massless theory for $\gmn$ and $\bmn$}\label{sec:action}

Our starting point will be the Einstein-Hilbert action for a metric $\gmn$ in the presence of an
antisymmetric tensor field $\bmn$, whose basic properties will be briefly reviewed in the following.
The action is,
\begin{equation}\label{gbaction}
S_{GB} = m_\mathrm{P}^2\int\dd^4x ~\sqrt{g} ~[R(g) - 2\Lambda] 
-  \frac1{2\cdot3!}\int\dd^4x~ \sqrt{g}~H_{\mu\nu\rho}H^{\mu\nu\rho}\,,
\end{equation}
where $H_{\mu\nu\rho}=3 \nabla_{[\mu}B_{\nu\rho]}
=\nabla_{\mu}B_{\nu\rho}+\nabla_{\rho}B_{\mu\nu}+\nabla_{\nu}B_{\rho\mu}$ 
are the components of the 3-form field strength. 
Here, $\nabla_{\mu}$ is the Levi-Civita connection of $\gmn$.
The field strength enters the action in the kinetic term for the massless field $\bmn$ in curved space, 
i.e.~all indices are raised with $g^{\mu\nu}$.
The Einstein-Hilbert action including the cosmological constant $\Lambda$ is the standard 
nonlinear kinetic term for a massless spin-2 field. 

The equations of motion obtained by varying the action with respect to the tensor
fields $\gmn$ and $\bmn$ are,
\begin{subequations}\label{gbeq}
\begin{align}
\mathcal{G}^{\mu\nu} &\equiv   R^{\mu\nu} -\frac12(R-2\Lambda)g^{\mu\nu} 
	- \frac{1}{4m_\mathrm{P}^2}(H^{\mu\rho\sigma}H^\nu{}_{\rho\sigma} - \frac16 H^2 g^{\mu\nu}) =0\,, \\
\mathcal{B}^{\mu\nu} &\equiv -\frac{1}{2m_\mathrm{P}^2}\nabla_\rho H^{\rho\mu\nu} =0\, .
\end{align}
\end{subequations}
Note that $\mathcal{G}_{\mu\nu}$ consists of the usual Einstein tensor 
$R_{\mu\nu}-\frac12g_{\mu\nu}R$
and the contribution from the ``matter source'' including $\bmn$ through its field strength. 
Also, $\mathcal{G}_{\mu\nu}$ is symmetric while $\mathcal{B}_{\mu\nu}$ is antisymmetric
in its two indices.


\paragraph{Local symmetries.}

The action in \eqref{gbaction} has several local symmetries (corresponding to gauge redundancies).
\begin{itemize}
\item[(i)] Diffeomorphisms which infinitesimally transform the coordinates as 
$\Delta_\xi x^\mu=\xi^{\mu}$, and the fields as,
\beqn\label{infdiff}
\Delta_\xi\gmn=\nabla_{\mu}\xi_{\nu}+\nabla_{\nu}\xi_{\mu}\,,
\qquad
\Delta_\xi\bmn=\xi^\rho\nabla_\rho\bmn+B_{\rho\nu}\nabla_\mu\xi^\rho-B_{\rho\mu}\nabla_\nu\xi^\rho\,.
\eeqn
It is well-known that these remove $4+4$ degrees of freedom, leaving 2 propagating modes in $\gmn$.
These correspond to the 2 helicity states of the massless spin-2 field.
\item[(ii)] $B$-gauge transformations under which only the $\bmn$ field transforms,
\beqn\label{btrafogen}
\Delta_\lambda\bmn=\nabla_\mu\lambda_\nu-\nabla_\nu\lambda_\mu\,,
\qquad
\Delta_\lambda\gmn=0\,.
\eeqn
It can be shown that these remove $3+2$ degrees of freedom, leaving 1 propagating mode in 
$\bmn$. This is of course consistent with the fact that a massless $\bmn$ field can be dualized to
a scalar in $D=4$.
\item[(iii)] Local Lorentz invariance, i.e.~SO(1,3) which transforms neither $\gmn$ nor $\bmn$ and 
therefore is not explicit in the tensor formulation. We see it only when we introduce a vierbein
$e^a_{~\mu}$ for the metric, $\gmn=e^a_{~\mu}\eta_{ab}e^b_{~\nu}$.
Then the infinitesimal local Lorentz transformations (LLT) read, 
\beqn
\Delta_\omega e^a_{~\mu}=\eta^{ab}\omega_{bc}e^c_{~\mu}\quad \text{with}~~\omega_{bc}=-\omega_{cb}\,,
\qquad
\Delta_\omega\bmn=\Delta_\omega\gmn=0\,.
\label{LorentzTrafos}
\eeqn
These precisely remove the 6 redundant degrees of freedom contained 
in the vierbein which do not enter the metric $\gmn$.
\end{itemize}
As a side remark, we also note that the diffeomorphisms in (\ref{infdiff}) 
can be written as a transformation of the 
vierbein $e^a_{~\mu}$ instead of the metric,
\beqn\label{vbdiff}
\Delta_\xi e^a_{~\mu}=\xi^\rho\nabla_\rho e^a_{~\mu} + e^a_{~\rho}\nabla_\mu\xi^\rho\,,
\qquad
\Delta_\xi\bmn=\xi^\rho\nabla_\rho\bmn+B_{\rho\nu}\nabla_\mu\xi^\rho-B_{\rho\mu}\nabla_\nu\xi^\rho\,.
\eeqn
The first term in the transformation of $e^a_{~\mu}$ drops out from the transformation of the metric
but is part of the proper transformation of a 1-form under diffeomorphisms.

\section{A new vierbein formulation}

Our aim is to package the $10+6=16$ degrees of freedom of $\gmn$ and $\bmn$ into
a single $4\times4$ matrix and identify the latter with the vierbein field $e^a_{~\mu}$.
As before, the metric is given by, 
\beqn\label{gdef}
g_{\mu\nu} = {e_\mu}^a\eta_{ab}{e^b}_\nu \,.
\eeqn 
In GR, 6 of the 16 components in $e^a_{~\mu}$ do not show up in $g_{\mu\nu}$ and are
thus unphysical due to local Lorentz invariance. 
Here we want them to remain physical and define the components of the field $\bmn$. 
This means that we cannot simply contract the Lorentz
index of $e^a_{~\mu}$ with SO(1,3) invariant tensors such as $\eta_{ab}$ or $\epsilon_{abcd}$
everywhere in the action.
We are forced to introduce a new object that carries a Lorentz index. Moreover, we need
to convert the Lorentz index of $e^a_{~\mu}$ into a coordinate index in order to translate its
components into those of the antisymmetric tensor. Therefore we choose the additional object 
to be a second vierbein $\te^a_{~\mu}$ which, when all its components are specified, 
introduces a fixed Lorentz frame. For instance, as a matrix, $\te^a_{~\mu}$ could just
be the identity $\delta^a_{~\mu}$.\footnote{The reader familiar with the 
topic will notice that this is the exact same situation as in massive gravity.
Ultimately, we would like to give dynamics to the frame field $\te^a_{~\mu}$
and promote it to a fully dynamical object. The resulting setup which is reminiscent of bimetric 
theory will be studied in~\cite{Markou:2018mhq}.}

Note that, by definition, there are no new dynamical degrees of freedom inside $\te^a_{~\mu}$.
On the other hand, the Lorentz components in $e^a_{~\mu}$ are no longer
pure gauge and potentially give rise to new dynamical degrees of freedom. 
This is dangerous, in general, because such new degrees of freedom are likely to give rise to ghost instabilities.
Below we will see that in our construction this problem is avoided. 
Our new theory will be build in such a way that the new propagating degrees of freedom
can be identified with the components of the antisymmetric tensor field $\bmn$.
The number of additional propagating degrees of freedom will be 2 in the massless and 3 in the massive theory.

Clearly there are infinitely many possibilities to build the tensor $\bmn$ out 
of the vierbeins $e^a_{~\mu}$ and $\te^a_{~\mu}$ but here we restrict to expressions
that are quadratic in the vierbeins. Then the
only antisymmetric rank-2 tensor without any free Lorentz indices is,
\beqn\label{bdef}
\bmn\equiv e^a_{~\mu}\eta_{ab}\te^b_{~\nu}- \te^a_{~\mu}\eta_{ab}e^b_{~\nu}\,.
\eeqn
Since the definitions of the tensors
cannot be directly inverted in a covariant way, it is not clear how exactly the 16
components split into the 10 components of $\gmn$ and 6 components of $\bmn$.
Thus, a priori, it seems that now $\gmn$ and $\bmn$ are no longer independent objects.

The next step is to insert the vierbein expressions for $g_{\mu\nu}$ and $\bmn$ 
into the action~\eqref{gbaction}.\footnote{Since, in our parametrization of the action, 
$\bmn$ has mass dimension 1 while $\gmn$ (and thus $e^a_{~\mu}$) have mass dimension 0, 
we take $\te^a_{~\mu}$ to have mass dimension 1. These conventions can of course easily be changed by 
performing a rescaling of the vierbeins.}
It is not clear that this will give a description equivalent to the tensor formulation
with independent $\gmn$ and $\bmn$. However, as we will show now, the equations of motion
for the vierbein guarantee the equivalence of the two formulations.

\subsection{Equations of motion}\label{sec:eom}

Using the chain rule it is easy to vary the action~\eqref{gbaction} with $\gmn$ and 
$\bmn$ given by \eqref{gdef} and \eqref{bdef} with respect to $e^a_{~\mu}$.
In this way we obtain the equation of motion for the vierbein,
\begin{equation}
\EE_a^{~\mu} \equiv\frac{\delta S}{\delta e^a_{~\mu}} = 
\frac{\delta g^{\rho \sigma}}{\delta e^a_{~\mu}}
\left. \frac{\delta S_{GB}}{\delta g^{\rho \sigma}}\right|_{B=\text{const.}} 
	+~  \frac{\delta B_{\rho \sigma}}{\delta e^a_{~\mu}} 
	\left.\frac{\delta S_B}{\delta B_{\rho \sigma}}\right|_{G=\text{const.}}  
	=0\,,
\end{equation}
where the variations of the tensors with respect to the vierbein are obtained from,
\begin{align}\label{vargbe}
\delta g_{\mu\nu} = \delta {e_\mu}^a\eta_{ab}e^b{}_\nu +  {e_\mu}^a\eta_{ab}\delta e^b{}_\nu 
\qquad
\delta B_{\mu\nu} = \delta {e_\mu}^a\eta_{ab}\te^b{}_\nu -  \te_\mu{}^a\eta_{ab}\delta e^b{}_\nu\,.
\end{align}
We then simply use the results for the tensor variations in \eqref{gbeq} to 
see that the equations of motion for the vierbein take the following form,
\begin{align}
\EE_a^{~\mu} &=
2\eta_{ab}{e_\nu}^b \Big[  R^{\mu\nu} -\frac12(R-2\Lambda)g^{\mu\nu} 
	- \frac{1}{4m_\mathrm{P}^2}(H^{\mu\rho\sigma}H^\nu{}_{\rho\sigma} - \frac16 H^2 g^{\mu\nu}) \Big] 
	-\frac{1}{m_\mathrm{P}^2}\eta_{ab} \te_\nu{}^b\nabla_\rho H^{\rho\mu\nu} 
	\nn\\
	&=0\,.
\end{align}
In terms of the tensors defined in \eqref{gbeq}, we can write these as,
\begin{equation}\label{EoMforE}
\EE_a^{~\mu} = 2\eta_{ab}{e^b}_\nu\mathcal{G}^{\mu\nu} + 2\eta_{ab} \te^b{}_\nu\mathcal{B}^{\mu\nu}=0\, .
\end{equation}
A priori, it is now not obvious that $\mathcal{G}^{\mu\nu}$ and $\mathcal{B}^{\mu\nu}$ should
vanish separately as they do in the tensor formulation of the theory.

\paragraph{Separating the equations.}
Consider the following antisymmetrized linear combination of the vierbein equations,
\begin{equation}\label{antisymmconstr}
2\eta^{ab}  e^{[\mu}{}_a \frac{\delta S}{\delta e^b{}_{\nu]}} 
= 2\eta^{ab} e^{[\mu}{}_a  \EE^{\nu]}_b = 0 .
\end{equation}
These correspond to a subset of 6 out of the 16 equations.
Inserting $\EE_a^{~\mu}$ from above and using the symmetry of $\mathcal{G}_{\mu\nu}$, 
they reduce to,
\begin{equation}\label{opP}
4 \te^a{}_\rho e^{[\mu}{}_a\mathcal{B}^{\nu]\rho} 
= 4 e^{[\mu}{}_a\te^{\nu]}{}_b \te^b{}_\sigma \te^a{}_\rho\mathcal{B}^{\sigma\rho} 
= - 2\tilde{P}^{\mu\nu}{}_{ab} \te^a{}_\rho\te^b{}_\sigma\mathcal{B}^{\rho\sigma} = 0
\end{equation}
where we have defined the operator $\tilde{P}^{\mu\nu}{}_{ab}\equiv 2e^{[\mu}{}_{[a}\te^{\nu]}{}_{b]}$.
As we will argue now, this operator is generically invertible, since
 on the space of antisymmetric matrices it acts just like a linear map. 
One way to see this is to introduce a combined index $M=[\mu\nu]$ 
and $A=[ab]$ for antisymmetrized pairs of normal indices. 
Then $\tilde{P}$ is like a matrix acting on a vector,
\begin{equation}
\tilde{P}^M_{~A}\mathcal{B}^A =0\,.
\end{equation}
Under the restriction to the space of antisymmetric matrices, $\tilde{P}^M_{~A}$ can 
thus generically be inverted.
Then the subset of equations in~\eqref{antisymmconstr} reduces to $\mathcal{B}^{\mu\nu}=0$, 
which are the equations for $B_{\mu\nu}$ in the tensor formulation. 
Upon inserting this back into the full equations \eqref{EoMforE}, they 
reduce to $\mathcal{G}_{\mu\nu} = 0$, which is the equation for $g_{\mu\nu}$ in the tensor formulation. 
This shows how the $\gmn$ and $\bmn$ equations separate and our vierbein formulation is
equivalent to the tensor formulation, provided that the operator $\tilde{P}^{\mu\nu}{}_{ab}$ is invertible.
More details on the independence of $\gmn$ and $\bmn$ and the relation to the invertibility of the operator 
can be found in appendix~\ref{app:independence}.

\subsection{Local symmetries}\label{sec:locsym}

We have explicitly broken local Lorentz invariance by making all 16 components in the vierbein 
physical.\footnote{In fact, local Lorentz invariance can be restored in two ways. As shown in
the previous subsection, $\gmn$ and $\bmn$ become independent dynamically. We can hence express $\gmn$ in
terms of a new vierbein whose components do not depend on $\bmn$ and whose Lorentz rotations thus leave the action
invariant. The other option is to make $\te^a_{~\mu}$ dynamical, in which case simultaneous Lorentz rotations
of the two vierbeins leave the action invariant. This latter case is particularly interesting and 
will be discussed in a future publication.}
The remaining local symmetries of the action~\eqref{gbaction} 
that we discussed in section~\ref{sec:action} must be recovered in the vierbein formulation. 
However, it is not immediately obvious how the diffeomorphisms and gauge transformations
can be written as a transformation of the vierbein $e^a_{~\mu}$ which now appears in both tensor fields.
In the following we discuss how exactly the symmetries emerge in this case.

\paragraph{Diffeomorphisms.}
Let us start from the diffeomorphisms transformation of the metric, 
$
\Delta_\xi\gmn =\nabla_{\mu}\xi_{\nu}+\nabla_{\nu}\xi_{\mu}\,.
$
As we already saw in \eqref{vbdiff}, this can be produced 
by the following transformation of the vierbein,
\beqn\label{Dediff}
\Delta_\xi e^a_{~\mu}=\xi^\rho\nabla_\rho e^a_{~\mu} + e^a_{~\rho}\nabla_\mu\xi^\rho+\eta^{ac}\omega_{cb}(\xi)e^b_{~\mu}\,.
\eeqn
Here we have allowed for a LLT with $\omega_{ab}=-\omega_{ba}$ 
depending on the diffeo parameter $\xi_\mu$, since this leaves the metric $\gmn$ invariant. 
Requiring the correct transformation behaviour for $\bmn$ will fix the form of $\omega_{ab}(\xi)$.
Namely, under diffeomorphisms, the 2-form should transform as,
\beqn\label{diffeosB}
\Delta_\xi\bmn&=&\xi^\rho\nabla_\rho\bmn+B_{\rho\nu}\nabla_\mu\xi^\rho-B_{\rho\mu}\nabla_\nu\xi^\rho\, \nn\\
 &=& \xi^\rho\nabla_\rho\bmn + 2\eta_{ab}(e^a_{~[\rho}\te^b_{~\nu]}\nabla_\mu\xi^\rho 
 - e^a_{~[\rho}\te^b_{~\mu]}\nabla_\nu\xi^\rho)\,.
\eeqn
On the other hand, inserting \eqref{Dediff} into the definition of $\bmn$, we obtain,
\beqn\label{should}
\Delta_\xi\bmn
&=&2\Delta_\xi e^a_{~[\mu} \eta_{ab}\te^b_{~\nu]}
=2\eta_{ab}\xi^\rho\nabla_\rho e^a_{~[\mu}\te^b_{~\nu]}
+2\eta_{ab}e^a_{~\rho}\te^b_{~[\nu}\nabla_{\mu]}\xi^\rho
+2e^a_{~[\mu}\te^b_{~\nu]}\omega_{ba}(\xi)\nn\\
&=&\xi^\rho\nabla_\rho\bmn - 2\eta_{ab}\xi^\rho\nabla_\rho \te^b_{~[\nu}e^a_{~\mu]}
+ \eta_{ab}e^a_{~\rho}(\te^b_{~\nu}\nabla_{\mu}\xi^\rho - \te^b_{~\mu}\nabla_{\nu}\xi^\rho)
-2e^{[a}_{~[\mu}\te^{b]}_{~\nu]}\omega_{ab}(\xi)\,. \nn\\
\eeqn
Comparing \eqref{should} to \eqref{diffeosB}, we see that we must have,
\beqn\label{defp}
- 2\eta_{ab}\xi^\rho\nabla_\rho \te^b_{~[\nu}e^a_{~\mu]} + \eta_{ab}\te^b_{~\rho}(e^a_{~\nu}\nabla_{\mu}\xi^\rho - e^a_{~\mu}\nabla_\nu\xi^\rho)
=2e^{[a}_{~[\mu}\te^{b]}_{~\nu]}\omega_{ab}(\xi)
\equiv P_{\mu\nu}{}^{ab}\omega_{ab}(\xi)
\eeqn
The object $P_{\mu\nu}{}^{ab}=2e^{[a}_{~[\mu}\te^{b]}_{~\nu]}$ 
is an operator similar to $\tilde{P}^{\mu\nu}{}_{ab}$
which we encountered earlier in eq.~\eqref{opP}. 
For the same reason as before, $P_{\mu\nu}{}^{ab}$ is generically invertible
on the space of antisymmetric matrices.
We thus find the local Lorentz transformation that is needed to generate the diffeomorphism of $\bmn$,
\begin{equation}
\omega_{ab}(\xi) =2(P^{-1})_{ab}{}^{\mu\nu}\Big(\xi^\rho\nabla_\rho\te^c_{~\mu}
+ \te^c_{~\rho}\nabla_{\mu}\xi^\rho\Big)\eta_{cd}e^d_{~\nu} \,,
\end{equation}
and the diffeomorphisms are generated by the following transformation of the vierbein,
\beqn\label{gtex}
\Delta_\xi e^a_{~\mu}=
\xi^\rho\nabla_\rho e^a_{~\mu}
+ e^a_{~\rho}\nabla_\mu\xi^\rho
+2\eta^{ac}\eta_{de}e^b_{~\mu}e^e_{~\nu}(P^{-1})_{cb}{}^{\sigma\nu}\Big(\xi^\rho\nabla_\rho\te^d_{~\sigma}
+ \te^d_{~\rho}\nabla_{\sigma}\xi^\rho\Big)\,.
\eeqn

\paragraph{Gauge transformations.}

We would also like to obtain the gauge transformation 
of the form $\Delta_\lambda B = \dd\lambda$ from a tranformation of $\Delta e^a_{~\mu}$. 
Since such a gauge transformation should leave $\gmn$ invariant, 
it must be recast as a local Lorentz transformation. At the linearized level 
it is thus given by,
\begin{equation}
\Delta_\lambda e^a{}_\mu = \omega^a{}_b(\lambda) e^b{}_\mu\,,\qquad \omega_{ab}(\lambda)=-\omega_{ba}(\lambda)\,.
\end{equation}
The form of $\omega_{ab}(\lambda)$ is now again determined by the correct transformation behaviour of $\bmn$.
Namely, the gauge transformation of the $B$-field gives,
\begin{equation}
\Delta_\lambda B_{\mu\nu} = 2\nabla_{[\mu}\lambda_{\nu]} 
 = 2 e_{[\mu}{}^{[a}\te_{\nu]}{}^{b]}\omega_{ab}(\lambda) 
 \equiv P_{\mu\nu}{}^{ab}\omega_{ab}(\lambda)\,.
\end{equation}
The object $P_{\mu\nu}{}^{ab}$  is the same operator as in equation~\eqref{defp}. 
We can again invert it and find the local Lorentz transformation corresponding 
to a given gauge transformation of $\bmn$,
\begin{equation}
\omega_{ab}(\lambda) = (P^{-1})_{ab}{}^{\mu\nu}\Delta B_{\mu\nu}
 =  2(P^{-1})_{ab}{}^{\mu\nu}\nabla_{[\mu}\lambda_{\nu]} .
\end{equation}
The transformation on the vierbein generating $B$-gauge transformations is thus,
\beqn\label{gtel}
\Delta_\lambda e^a{}_\mu=2\eta^{ac}(P^{-1})_{cb}{}^{\rho\sigma}\nabla_{[\rho}\lambda_{\sigma]} e^b{}_\mu\,.
\eeqn

\paragraph{Counting degrees of freedom.}
It is worth commenting on the degree of freedom counting at this point.
We start out with the 16 components in $e^a_{~\mu}$. The gauge symmetry in
eq.~(\ref{gtex}) resembles a diffeomorphism plus a local Lorentz transformation
and depends on 4 gauge parameter components $\xi^\mu$ in total.  
In the separated equation $\mathcal{G}_{\mu\nu}=0$ for $\gmn$, 
it removes $4+4$ components, leaving us with 2 propagating degrees of freedom in $\gmn$.
The gauge symmetry for $\bmn$ in eq.~(\ref{gtel}) resembles a local Lorentz transformations and does
not affect $\gmn$. It depends on 3 gauge parameter components in the divergence free vector $\lambda_\mu$.
In the separated equation $\mathcal{B}_{\mu\nu}=0$ for $\bmn$, it removes $3+2$ components,
leaving us with 1 propagating degree of freedom in $\bmn$.

Since the equations separate and the gauge symmetries can be reproduced, we 
conclude that the vierbein formulation for the massless theory
is entirely equivalent to the tensor formulation in section~\ref{sec:action}.

\section{The massive theory}\label{sec:addingmass}

We would like to add interactions to the massless theory which render the fields massive.
In the tensor formulation with independent $\gmn$ and $\bmn$, 
one option would be to include the standard mass term for the antisymmetric tensor
which is of the form,
\beqn\label{standardb}
-m^2\int\dd^4x~\sqrt{g}~ B_{\mu\nu}B_{\rho\sigma}g^{\mu\rho}g^{\nu\sigma}\,.
\eeqn
This gives a consistent theory, describing a massive antisymmetric tensor in curved background
given by a metric which remains massless.
We perform a 3+1 split and degree of freedom counting for the corresponding action 
in appendix~\ref{admmassb}.
A massive $\bmn$ field has 3 propagating degrees of freedom, which is consistent with the fact 
that in $D=4$ it is dual to a massive vector (while a massless $\bmn$ is dual to 
a massless scalar)~\cite{Smailagic:2001ch}.

Here we are going to take a different approach and show that a mass term that is usually used 
for $\gmn$ can make both fields $\gmn$ and $\bmn$ massive. This is achieved precisely by our 
identification of the field $\bmn$ with the 6 additional components in the vierbein $e^a_{~\mu}$.
In section~\ref{sec:ADM} we show explicitly that the resulting action has $5+3$ propagating modes,
as expected for a massive $\gmn$ and a massive $\bmn$. 
No additional (potentially ghost-like) degrees of freedom
are introduced.\footnote{It turns out that it is then inconsistent to also add the  
standard mass term for $\bmn$ in \eqref{standardb} to this setup.} However,
the fluctuation of the $\bmn$ field around maximally backgrounds turns out to be tachyonic, 
as we will see below.

In order to set the stage, we begin by reviewing the mass term for the metric $\gmn$ which
usually does not contain the $\bmn$ field, since its consistency requires $\bmn=0$.

\subsection{Massive gravity in vierbein formulation}

Massive gravity extends the Einstein-Hilbert term of GR by a mass potential which assumes 
a rather simple form when written in terms of the vierbein field $e^a_{~\mu}$. The only nontrivial 
term that can be built from this vierbein alone is $\det e$, which corresponds to
the gauge-invariant cosmological constant term and is not part of the mass giving potential. 
Writing down the mass term therefore requires introducing a second vierbein 
$\te^a_{~\mu}$. This field can be made dynamical in a consistent way 
but for now we focus on the case with non-dynamical~$\te^a_{~\mu}$.

The ghost-free action for massive gravity then is,
\begin{align}\label{MG}
S_\mathrm{MG}=
m_\mathrm{P}^2\int\dd^4 x\,\sqrt{g}\,\big(R(g)-\Lambda\big)
- m_\mathrm{P}^2m^2\int \epsilon_{abcd}\big(
&b_1 \,e^a\wedge e^b \wedge e^c \wedge \te^d
+b_2 \,e^a\wedge e^b \wedge \te^c \wedge \te^d
\nn\\
 \hspace{70pt}+~
&b_3\, e^a\wedge \te^b \wedge \te^c \wedge \te^d\big)\,,
\end{align}
where the interaction parameters $b_n$ are arbitrary constants.
As advertised, the action contains an Einstein-Hilbert kinetic term for the metric $\gmn=e^a_{~\mu}\eta_{ab}e^b_{~\nu}$
and an interaction potential giving mass to the spin-2 field. The interactions are forced to be of the above
totally antisymmetrized form in order to guarantee the absence of the Boulware-Deser ghost~\cite{Boulware:1973my}. 
The Hamiltonian analysis for the action \eqref{MG} has been
performed in Ref.~\cite{Hassan:2011hr, Hassan:2011tf, Hinterbichler:2012cn}. The mass term can be
consistently added to \eqref{gbaction} with independent tensor fields,
in which case the spin-2 degrees of freedom become massive while
$\bmn$ remains massless.

The equations of motion for the vierbein field $e^a_{~\mu}$ following from \eqref{MG} 
contain a set of constraints arising from the potential. These read~\cite{Hinterbichler:2012cn},
\beqn\label{LLTconstr}
\eta^{ab}  e^{[\mu}{}_a \frac{\delta S_\mathrm{MG}}{\delta e^b{}_{\nu]}} =0\,.
\eeqn
The kinetic term does not contribute to this antisymmetric part of the equations
due to its invariance under local Lorentz transformations, $\Delta e^a_{~\mu}=\eta^{ab}\omega_{bc} e^c_{~\mu}$
with $\omega_{bc}=-\omega_{cb}$.
The constraints \eqref{LLTconstr} have a unique covariant solution which reads,
\beqn\label{LLTconstrsol}
e^a_{~\mu}\eta_{ab}\te^b_{~\nu}-\te^a_{~\mu}\eta_{ab}e^b_{~\nu}=0\,.
\eeqn
These 6 constraints determine the extra components in $e^a_{~\mu}$ which do not show up in $\gmn$.
In fact, it is crucial that \eqref{LLTconstrsol} holds because it ensures the absence of the Boulware-Deser
ghost in the action~\cite{deRham:2015cha}. 
It also allows us to express the vierbein potential in terms of the respective metrics
$\gmn$ and $\fmn=\te^a_{~\mu}\eta_{ab}\te^b_{~\nu}$, in which case the interactions 
depend on the square-root matrix $\sqrt{g^{-1}f}$.

Massive gravity does therefore not allow configurations with 
$e^a_{~\mu}\eta_{ab}\te^b_{~\nu}-\te^a_{~\mu}\eta_{ab}e^b_{~\nu}\neq0$.\footnote{In fact, the  
antisymmetric combination of vierbeins can be non-vanishing only in theories involving at least
3 vierbein fields and a very particular type of ghost-free coupling~\cite{Hassan:2018mcw}.}
However, this result crucially depends on the fact that  
$\bmn\equiv e^a_{~\mu}\eta_{ab}\te^b_{~\nu}-\te^a_{~\mu}\eta_{ab}e^b_{~\nu}$
does not possess a kinetic term.
In section \ref{sec:ADM} we will demonstrate that by giving dynamics to the antisymmetric 
components we can have $\bmn\neq0$ without introducing the Boulware-Deser ghost.

\subsection{The vierbein action with mass term}\label{sec:newmass}

We now add the ghost-free mass term for the vierbein field $e^a_{~\mu}$ in \eqref{MG} 
to the action~\eqref{gbaction}, identifying the non-dynamical vierbein $\te^a_{~\mu}$
in the antisymmetric tensor with the reference field in the mass term.\footnote{Using two different reference
fields would not change the conclusions below but seems like an unnecessary complication of our minimal setup.}
The full action $S_\mathrm{m} = S_{GB} + S_V$ reads,
\beqn\label{gbactionmass}
S_\mathrm{m}
&=& m_\mathrm{P}^2\int\dd^4x ~\sqrt{g} ~[R(g) - 2\Lambda] 
-  \frac1{2\cdot3!}\int\dd^4x~ \sqrt{g}~H_{\mu\nu\rho}H^{\mu\nu\rho}
\nn\\
&-&m_\mathrm{P}^2m^2\int \epsilon_{abcd}
\Big(
b_1 ~e^a\wedge e^b \wedge e^c \wedge \te^d
+b_2 ~e^a\wedge e^b \wedge \te^c \wedge \te^d
+ b_3~ e^a\wedge \te^b \wedge \te^c \wedge \te^d\Big)\,,
\nn\\
\eeqn
with $\gmn$ and $\bmn$ defined in terms of the vierbein as in \eqref{gdef} and \eqref{bdef}.\footnote{Note that since in our 
parametrization, the vierbein $\te^a_{~\mu}$
has mass dimension 1, the $b_n$ parameters have mass dimension $-n$.}
The mass term breaks both the diffeomorphism and the gauge symmetry of the $\bmn$ field,
which were present in the massless theory.
We will show that the above action
provides a nonlinear description of interacting massive $\gmn$ and massive $\bmn$.

In contrast to massive gravity, since now $\bmn\neq0$, 
the interaction potential does not possess a formulation in terms of metrics. 
Nevertheless, it is still possible to separate the equations of motion into 
an antisymmetric and symmetric part, which contain exactly the same kinetic structures as 
in the massless theory and its equivalent tensor formulation.

\paragraph{Separating the equations.}
The mass term gives an extra contribution to the equation of motion of $e^a_{~\mu}$,
which we shall denote by
$\mathcal{V}_a^{~\mu}\equiv -{(m_\mathrm{P}^2 \det e)}^{-1}\frac{\delta S_V}{\delta e^a_{~\mu}}$. 
The full equations -- extending \eqref{EoMforE} -- then read,
\begin{equation}\label{fulleq}
\EE_a^{~\mu} = 2\eta_{ab}{e^b}_\nu\mathcal{G}^{\mu\nu} + 2\eta_{ab} \te^b{}_\nu\mathcal{B}^{\mu\nu} 
+ \mathcal{V}_a^{~\mu}=0\, ,
\end{equation}
where $\mathcal{G}^{\mu\nu}$ and $\mathcal{B}^{\mu\nu}$ are still the same as in \eqref{gbeq}.
After multiplying by $\frac{1}{2}\eta^{ac}{e_c}^\rho$ and 
taking the antisymmetric part of the resulting equation, 
a few simple manipulations lead to,
\beqn\label{massbeq}
\mathcal{B}^{\mu\nu}- (\tilde{P}^{-1})_{\rho\sigma}{}^{ab}\te^{\mu}_{~a}\te^{\nu}_{~b}
e^{\rho}_{~c}\eta^{cd}\mathcal{V}^{~\sigma}_d=0 \,,
\eeqn
instead of the earlier result $\mathcal{B}^{\mu\nu}=0$. Here, 
$\tilde{P}^{\mu\nu}{}_{ab}\equiv 2e^{[\mu}{}_{[a}\te^{\nu]}{}_{b]}$
is the same invertible operator that we encountered in the massless case.
Plugging \eqref{massbeq} back into the full equation \eqref{fulleq}, we find,
\beqn\label{massgeq}
\mathcal{G}^{\mu\nu} +  
(\tilde{P}^{-1})_{\rho\sigma}{}^{ab}\te^{\mu}_{~a}e^{\nu}_{~b}
e^{\rho}_{~c}\eta^{cd}\mathcal{V}^{~\sigma}_d
        + \frac12 e^\nu_{~a}\eta^{ab}\mathcal{V}^{~\mu}_b = 0 \, .
\eeqn 
By construction, the antisymmetric part of this expression identically vanishes\footnote{This
is easy to verify directly using 
$(\tilde{P}^{-1})_{\rho\sigma}{}^{ab}e^{[\nu}_{~a}\te^{\mu]}_{~b}=
\frac12(\tilde{P}^{-1})_{\rho\sigma}{}^{ab}\tilde{P}_{ab}{}^{\nu\mu}
=\frac12\delta^{[\nu}_{~[\rho}\delta^{\mu]}_{~\sigma]}$.}
and we are left with a symmetric equation for $\mathcal{G}^{\mu\nu}$.
It captures the contribution from the interaction term to the equation of motion for the components
in the metric $g_{\mu\nu}$. From the form of the equations \eqref{massbeq} and \eqref{massgeq} 
we conclude that the kinetic structures have not been altered with respect to the massless theory.

We also observe that the action \eqref{gbactionmass} in fact contains ghost-free massive gravity 
in the form of a particular solution to \eqref{massbeq}. 
This solution corresponds to $\bmn=0$, in which case \eqref{massgeq}
reduces to the equation of motion for the vierbein following from varying the massive
gravity action in \eqref{MG}. This shows that our new theory can be viewed as a generalization of 
massive gravity.

\subsection{Linear theory}

In this section we study the spectrum of linear perturbations around the 
background $\bar{e}^a_{~\mu}$ for which the dynamical vierbein is proportional to the reference frame,
$\te^a_{~\mu}=c \bar{e}^a_{~\mu}$ with constant $c\neq 0$ of mass dimension 1. 
Clearly, for such a background we have that $\bar{B}_{\mu\nu}=0$,
which solves eq.~\eqref{massbeq}. Plugging the proportional ansatz into \eqref{massgeq},
we find that $\bar{g}_{\mu\nu}$ is the metric of a maximally symmetric spacetime with 
$R_{\mu\nu}(\bar{g})=\Lambda_g\bar{g}_{\mu\nu}$. The constant background curvature is given by
the following combination of parameters,
\beqn
\Lambda_g=\Lambda +3m^2\Big( 3  b_1c  + 2 b_2c^2 +  b_3c^3 \Big)\,.
\eeqn
Next, we write the nonlinear vierbein in terms of this background and a small fluctuation,
\beqn
e^a_{~\mu}=\bar{e}^a_{~\mu}+\delta e^a_{~\mu}\,.
\eeqn
We split the fluctuation into a symmetric and an antisymmetric component as follows,
\beqn
\eta_{ab}\be^b_{~\nu}\delta e^a_{~\mu}\equiv h_{\mu\nu}+b_{\mu\nu}\,,
\eeqn
where,
\beqn
h_{\mu\nu}=h_{\nu\mu}=\eta_{ab}\be^b_{~(\nu}\delta e^a_{~\mu)}\,,\qquad
b_{\mu\nu}=-b_{\nu\mu}=\eta_{ab}\be^b_{~[\nu}\delta e^a_{~\mu]}\,.
\eeqn
Due to this definition, the fluctuation of the metric is $\delta\gmn = 2h_{\mu\nu}$ and the 
fluctuation of the $B$-field is $\delta\bmn = 2cb_{\mu\nu}$. By a tedious but straightforward computation 
one can show that the linearized version of the full equations of motion \eqref{fulleq} 
can be separated and diagonalized into the two sets,
\begin{subequations}
\beqn
\mathcal{E}_{\mu\nu}^{~~\rho\sigma}h_{\rho\sigma}
-\Lambda_g\big(h_{\mu\nu}-\tfrac{1}{2}h_{\rho\sigma}\bar{g}^{\rho\sigma}\bar{g}_{\mu\nu}\big)
-\tfrac{m_\mathrm{FP}^2}{2}\big(h_{\mu\nu}-h_{\rho\sigma}\bar{g}^{\rho\sigma}\bar{g}_{\mu\nu}\big)&=&0\,,
\\
\bar{\nabla}^\rho\bar{\nabla}_{[\rho} b_{\mu\nu]}-m_b^2b_{\mu\nu}&=&0\,.
\eeqn
\end{subequations}
Here we have defined the linearized Einstein tensor 
in terms of the covariant derivative $\bar{\nabla}_\mu$ compatible with the
background metric $\bar{g}_{\mu\nu}$, 
\begin{align}\label{kinopds}
{\mathcal{E}}^{~~\rho\sigma}_{\mu\nu}h_{\rho\sigma} 
=-\tfrac{1}{2}\Big[\delta^\rho_\mu\delta^\sigma_\nu\bar{\nabla}^2
+\bar g^{\rho\sigma}\bar{\nabla}_\mu\bar{\nabla}_\nu 
&-\delta^\rho_\mu\bar{\nabla}^\sigma\bar{\nabla}_\nu
-\delta^\rho_\nu\bar{\nabla}^\sigma\bar{\nabla}_\mu \nn\\
&-\bar{g}_{\mu\nu}\bar g^{\rho\sigma}\bar{\nabla}^2 
+\bar{g}_{\mu\nu}\bar{\nabla}^\rho\bar{\nabla}^\sigma\Big]h_{\rho\sigma}\,.
\end{align}
These are the linear equations for a massive spin-2 field and a massive antisymmetric tensor
in maximally symmetric backgrounds. Their masses are given by the following parameter 
combinations,
\begin{subequations}
\beqn
m_\mathrm{FP}^2&=&m^2\Big( 3 b_1c  + 4 b_2c^2 + 3 b_3c^3 \Big)\,,\\
m_b^2&=& -\frac{m_\mathrm{P}^2}{3c^2} m^2\Big( 3 b_1c  + 4 b_2c^2 + 3 b_3c^3 \Big) \,.
\eeqn
\end{subequations}
Recall that $c$ has mass dimension 1 and $b_n$ has mass dimension $-n$, such that
$m_\mathrm{FP}$ and $m_b$ have the correct dimension.  We note that that these two masses are related by 
$m_b^2 =- \frac{m_\mathrm{P}^2}{3c^2}m_\mathrm{FP}^2$. Since $c$ is an arbitrary parameter in our setup, 
the two masses are independent. For $c^2>0$ and our sign choice in front of the vierbein interaction
potential, the antisymmetric fluctuation has a negative squared mass and therefore is a 
tachyon.\footnote{We are grateful to James Bonifacio for bringing this to our attention.} This
means that the vacuum solution with $\te^a_{~\mu}=c \bar{e}^a_{~\mu}$ is unstable (at least for real values
of $c$).
 

\section{ADM analysis}\label{sec:ADM}

In this section we will perform a 3+1 split for the vierbein theory defined by the
action~\eqref{gbactionmass}. This will allow us to count the number of 
propagating degrees of freedom at the nonlinear level and confirm that it agrees with
the linear theory.

We parameterize the general vierbein $e^a_{~\mu}$ as a Lorentz transformation of a
gauge-fixed vierbein~$E^a_{~\mu}$,
\beqn\label{vbADM}
e^a_{~\mu}&=&\Lambda^a_{~b}E^b_{~\mu}
=\begin{pmatrix}
\Gamma & \Gamma v_\beta \\
\Gamma v^\alpha & \mathcal{V}^\alpha_{~\beta}
\end{pmatrix}
\begin{pmatrix}
N & 0\\
E^\beta_{~j}N^j & E^\beta_{~i}
\end{pmatrix}\,,
\eeqn
where
\beqn
\Gamma&\equiv& \frac{1}{\sqrt{1-v^\alpha v_\alpha}}\,,\qquad 
\mathcal{V}^\alpha_{~\beta}~\equiv ~\delta^\alpha_{~\beta}+\frac{\Gamma^2}{1+\Gamma}v^\alpha v_{\beta}\,.
\eeqn
Here $\alpha,\beta=1,2,3$ are spatial Lorentz and $i,j=1,2,3$ are spatial coordinate indices.
The Lorentz rotations sit entirely in $E^\beta_{~i}$, that is,
\beqn
E^\beta_{~i}=R^\beta_{~\alpha}\bar{E}^\alpha_{~i}\,,
\eeqn
for some gauge-fixed $\bar{E}^\alpha_{~i}$. The gauge-fixed vierbein depends on 10 independent 
field variables and the Lorentz matrix contains the 6 additional components that 
do not show up in the kinetic term for $\gmn$, but only appear in the mass potential.

For simplicity, let us take $\te^a_{~\mu}=\delta^a_{~\mu}$.\footnote{For notational simplicity, 
we neglect the fact that $\te^a_{~\mu}$ is of mass dimension 1, since scales are irrelevant for the 
ADM analysis.}
The expressions with a general form for $\te^a_{~\mu}$ can be found in appendix~\ref{app:adm}
and all our results derived in the following straightforwardly generalize to that case. 
With this simple choice for $\te^a_{~\mu}$ we have that,
\beqn
\te^a_{~\mu}\eta_{ab}e^b_{~\nu}
=
\begin{pmatrix}
-N\Gamma -\Gamma v_\alpha E^\alpha_{~k}N^k ~~&~~ -\Gamma v_\alpha E^\alpha_{~j}\\&\\
 N\Gamma \delta_{i\alpha} v^\alpha +\delta_{i\alpha} \mathcal{V}^\alpha_{~\beta}E^\beta_{~k} N^k  
 ~~&~~ \delta_{i\alpha} \mathcal{V}^\alpha_{~\beta}E^\beta_{~j}
\end{pmatrix}\,,
\eeqn
from where we read off the components of the antisymmetric tensor,
\begin{subequations}
\beqn
B_{ij}&=&2\delta_{\alpha[i} E^\beta_{~j]}\mathcal{V}^\alpha_{~\beta}
=2\delta_{\alpha[i} \bar{E}^\beta_{~j]}\mathcal{V}^\alpha_{~\epsilon} R^{\epsilon}_{~\beta}\,,\\
B_{0i}
&=&\Gamma v_\alpha R^{\alpha}_{~\beta}\bar{E}^\beta_{~i} 
+N\Gamma \delta_{i\alpha} v^\alpha +\delta_{i\alpha} \mathcal{V}^\alpha_{~\beta}
R^{\beta}_{~\epsilon}E^\epsilon_{~k} N^k  \,.
\eeqn
\end{subequations}
The expressions for the Einstein-Hilbert action in terms of ADM variables are 
well-known~\cite{Arnowitt:1962hi} and those for the mass potential were 
discussed in detail in Ref.~\cite{Hinterbichler:2012cn, deRham:2015cha}. 
These results are summarized in appendix~\ref{app:adm}, where
we also derive the expressions for the $\bmn$ kinetic term.
From those results it follows that, before solving any of the constraints, 
the action assumes the following form,
\beqn\label{almfinadmact}
S=\int\dd^4x\Big(\bar{\Pi}_\alpha^{~j}\dot{\bar{E}}^\alpha_{~j}+\Pi^{ij}\dot{B}_{ij} +N\mathcal{C} +N^i\mathcal{C}_i
+\mathcal{L}_\mathrm{rest}\Big)\,,
\eeqn
where $\bar{\Pi}_\alpha^{~j}$ and $\Pi^{ij}$ are the canonical conjugate momenta of ${\bar{E}}^\alpha_{~j}$ and 
${B}_{ij}$, respectively. The canonical momenta for the remaining variables ($N$, $N^i$, $B_{0j}$
and the Lorentz fields $v^\alpha$, $R^\alpha_{~\beta}$) all vanish and hence the purely spatial 
components ${\bar{E}}^\alpha_{~j}$ and ${B}_{ij}$ are the only dynamical variables.
Crucially, the constraints $\mathcal{C}$ and $\mathcal{C}_i$ as well as the additional terms in 
$\mathcal{L}_\mathrm{rest}$ do not contain the lapse $N$ nor the shift components $N^i$ of the vierbein. 

The next step is to perform an invertible field redefinition for the 
3 independent components of the Lorentz rotation 
$R^{\alpha}_{~\beta}$.\footnote{We can extract the 3 independent components of $R^{\alpha}_{~\beta}$ 
explicitly, for instance by using the Caley transform for orthogonal tranformations, 
$R = (\eta+A)^{-1}(\eta-A)$ with $A^\mathrm{T}=-A$ which has 3 independent entries~\cite{Caley}.} 
Namely, we introduce
$R'_{ij}\equiv \delta_{\alpha i} \bar{E}^\beta_{~j}\mathcal{V}^\alpha_{~\epsilon} R^{\epsilon}_{~\beta}$
whose 3 independent components now replace the 3 rotation parameters.
Note that this redefinition does not involve the lapse and shift functions.
The spatial components ${B}_{ij}$ are precisely the antisymmetric part of $R'_{ij}$.
We can thus interpret the 3 independent components of ${B}_{ij}=R'_{[ij]}$ 
as giving dynamics to the 3 components of $R^{\alpha}_{~\beta}$. This implies that,
in contrast to massive gravity, the equations of motion for the rotations $R^{\alpha}_{~\beta}$ 
are now dynamical equations instead of constraints.

The set of nondynamical fields consists of $N$, $N^i$ and the boosts $v^\alpha$.
Their equations are therefore constraints, which take the following form,
\begin{subequations}
\beqn
\frac{\delta S}{\delta N}=\mathcal{C}(v, \bar{E}, R')=0\,,\\
\frac{\delta S}{\delta N^i}=\mathcal{C}_i(v, \bar{E}, R')=0\,,\label{niconst}\\
\frac{\delta S}{\delta  v^\alpha}
=N\frac{\delta \mathcal{C}}{ \delta v^\alpha} 
+N^i\frac{\delta \mathcal{C}_i}{\delta v^\alpha}
+\frac{\delta \mathcal{L}_\mathrm{rest}}{\delta v^\alpha}=0
\,.\label{vconst}
\eeqn
\end{subequations}
We can solve the 3 equations in \eqref{niconst} for the 3 components of the boost vector,
yielding a solution $v^\alpha=v^\alpha(\bar{E}, R)$, which is independent of $N$ and $N^i$. 
Then we solve the 3 equations in \eqref{vconst} for the shift components $N^i$. 
Since these equations are linear in $N$ and $N^i$, they result 
in a solution of the form,
\beqn
N^i=A^i\big(v(\bar{E}, R'\big),\bar{E}, R')+ N B^i\big(v(\bar{E}, R'),\bar{E}, R'\big)\,,
\eeqn
which is linear in $N$. Thus, plugging the solutions to \eqref{niconst}  
and \eqref{vconst} for $v^\alpha$ and $N^i$  back into the action, we arrive at the following form,
\beqn\label{finadmact}
S=\int\dd^4x\Big(\bar{\Pi}_\alpha^{~j}\dot{\bar{E}}^\alpha_{~j}+\Pi^{ij}R'_{[ij]} +N\mathcal{C}'(\bar{E}, R')
+\mathcal{L}'_\mathrm{rest}(\bar{E}, R')\Big)\,.
\eeqn
The action now contains $6+3$ dynamical variables in ${\bar{E}}^\alpha_{~j}$ and ${B}_{ij}$
which are subject to one constraint coming from the $N$ equation of motion, $\mathcal{C}'=0$.
This means we have $8=5+3$ propagating degrees of freedom, in agreement with the linear theory 
for a massive spin-2 and a massive $B$-field.

Let us emphasize again that for massive gravity (i.e.~in the absence of a kinetic term for $\bmn$), 
it was shown that the only ghost-free configuration is $\bmn=0$~\cite{deRham:2015cha}. 
More precisely, in massive gravity the equations for the non-dynamical rotation matrices 
$R^\alpha_{~\beta}$ are constraints which have to be solved in order to determine whether the final
action is linear in the lapse $N$. If the algebraic solution for $R^\alpha_{~\beta}$ was such that 
the antisymmetric tensor was non-vanishing, $\bmn\neq 0$, 
the additional constraint needed to remove the Boulware-Deser ghost would be lost. 
The reason why our setup does not re-introduce the Boulware-Deser ghost, even though 
it contains $\bmn\neq0$, is the dynamical nature of the antisymmetric tensor. 
We do not need to solve the $R^\alpha_{~\beta}$ equations in order to see the full set 
of constraints. As a consequence, our theory contains the 3 propagating modes of the massive
$\bmn$ field, which all appear with a standard (ghost-free) kinetic term.

\section{Discussion}

We have shown that the new classical vierbein theory defined by the action \eqref{gbactionmass}
is a ghost-free nonlinear theory for a massive spin-2 and a massive antisymmetric tensor
field. Let us stress once more that our entire discussion was restricted to $D=4$ 
but the results straightforwardly generalize to any spacetime dimension. 
The physical vierbein $e^a_{~\mu}$ of the space time metric also defines the antisymmetric
tensor whose kinetic term gives dynamics to the Lorentz components in $e^a_{~\mu}$.
Since $\bmn\neq 0$ in general, the action is not expressible in terms of tensor fields. 

The new action can be viewed as a deformation of the massless theory for $\gmn$ and $\bmn$.
At the same time, it is a non-trivial generalization of ghost-free massive gravity, which 
is in fact contained in the theory and corresponds to the solution $\bmn=0$. However, this
vacuum is unstable due to the presence of the tachyonic mass pole for the antisymmetric tensor fluctuation.
Our results can be extended to the case where the reference frame $\te^a_{~\mu}$ is dynamical,
or even to a setup with multiple dynamical vierbein fields. This yields generalizations 
of bi- and multimetric gravity which are manifestly invariant under local Lorentz transformations
and diffeomorphisms. We will discuss this in detail in a future publication~\cite{Markou:2018mhq}.
The setup studied in this work can be obtained from the fully dynamical action by freezing
out the dynamics of $\te^a_{~\mu}$. This corresponds to the exact same parameter limit in which 
massive gravity is obtained from bimetric theory~\cite{Baccetti:2012bk, Hassan:2014vja}. 

As a ghost-free theory with $\bmn=0$, our result could also allow for the inclusion of
new matter couplings into multi-vierbein theories. Ref.~\cite{Hinterbichler:2015yaa} suggested to couple
a linear combination of vierbein fields to the matter sector. This construction failed to
be ghost-free precisely because the antisymmetric vierbein combination was non-vanishing but
also non-dynamical~\cite{deRham:2015cha}. With a kinetic term for $\bmn$, this problem could be
resolved.

Our ADM analysis shows that the action propagates the correct number of degrees of freedom
at the nonlinear level. In particular, the Boulware-Deser ghost (which is a notoric problem
in theories with massive spin-2 fields) is removed from the physical spectrum by an additional
constraint. However, the mass spectrum around maximally symmetric solutions contains a
tachyon. Determining whether the full Hamiltonian (which can be directly obtained from \eqref{finadmact}) 
is bounded from below requires explicitly solving all constraints, which is beyond the scope of this work.
In principle there could be more ghosts or further pathologies hidden 
in the theory, as part of the dynamical modes in $\gmn$ and $\bmn$. We believe that this is 
rather unlikely because, as we showed in sections~\ref{sec:eom} and~\ref{sec:newmass}, 
the kinetic structures of the massless theory for independent tensor fields are not modified 
by the identification of $\bmn$ with the antisymmetric vierbein components nor by the presence 
of the mass term.  

The tachyonic instability on constant curvature backgrounds may be worrisome but can 
be cured in various ways. First, in 4-dimensional Anti-de-Sitter spacetime (i.e.~for $\Lambda<0$), the mass pole
remains unitary above the Breitenlohner-Freedman bound~\cite{Breitenlohner:1982bm}.  
For spin-2 fields, this bound reads $m^2_{\mathrm{FP}}>\frac{9}{4}\Lambda$. Swapping the overall
sign of the mass potential, the mass pole for the $b_{\mu\nu}$ fluctuation becomes unitary.
If we then in addition restrict the parameters in the action in order to satisfy the bound for $m^2_{\mathrm{FP}}$,
also the spin-2 mass pole is unitary. Second, we could instead imagine swapping the overall sign of the 
kinetic term for $\bmn$. This would turn the fluctuation into a tachyonic ghost. Now a field redefinition
$b_{\mu\nu}\rightarrow i b_{\mu\nu}$ would result in a unitary mode. Such a redefinition may have interesting 
consequences on the structure of the nonlinear theory. Investigating this possibility is beyond the scope of
this work.

Another potential issue is the question of causality. In Ref.~\cite{Hassan:2017ugh} it was shown 
that the possibility to address the question of causal propagation in ghost-free bimetric theory 
requires precisely the vanishing of the antisymmetric vierbein combination. The presence of 
our dynamical $\bmn$ could in principle spoil this feature, which may be indicated by the tachyonic 
instability that we encountered. 
A different approach to studying the causal structure of massive spin-2 theories 
in the framework of scattering amplitudes was taken in Ref.~\cite{Hinterbichler:2017qyt,
Bonifacio:2017nnt,deRham:2017zjm,deRham:2017xox,deRham:2018qqo}. 
A thorough investigation of this issue is beyond the scope of this paper. 

It would be very interesting to develop a geometrical interpretation for our setup. 
The vierbein (or frame field) $e^a_{~\mu}$ together with the invariant 
$\eta_{ab}$ of the SO(1,3) Lorentz structure group define the metric 
$\gmn=e^a_{~\mu}\eta_{ab} e^b_{~\nu}$ of the pseudo-Riemannian geometry. 
For certain configurations, the presence of the $\bmn$ field can be interpreted as 
an additional symplectic structure on the manifold with the structure group Sp(4). 
This can be seen by introducing a symplectic frame field $b^a_{~\mu}$ which, together with the 
symplectic invariant $\Omega_{ab}$, gives a decomposition $\bmn = b^a_{~\mu}\Omega_{ab} b^b_{~\nu}$. 
It is clear that then $\bmn$ is invariant under symplectic transformations of its frame field. 
Well-studied cases with \textit{compatible} metric and symplectic structure 
are manifolds with (almost) Hermitian or (almost) K\"ahler geometry. 
In our setup, $\gmn$ and $\bmn$ are intimately related via the vierbein $e^a_{~\mu}$ 
and it could be interesting to investigate the consequences when their respective structures are 
made compatible with each other. In spirit this would be similar to recent attempts to construct the 
geometry of quantum gravity~\cite{Freidel:2017yuv, Freidel:2018tkj}.  
Moreover our setup may allow to relate the symplectic interpretation of the $\bmn$ field 
to that of a geometry with torsion. The torsion 
perspective might also be related to teleparallel theories where the antisymmetric 
part of the linearized vierbein becomes dynamical, see section 4.6 of \cite{Ortin:2004ms}. Moreover, 
it would be interesting to see whether there is a relation to the non-commutative setup of Ref.~\cite{deCesare:2018cjr}.

As outlined in the introduction, in string theory the antisymmetric tensor field appears
together with the metric as massless excitation of the closed string. In the low-energy effective
theory, together with the dilaton field, they make up the bosonic sector of supergravity. 
It would thus be interesting to investigate the consequences of adding our mass term to the 
supergravity action, which could also be used to deform the structure of double field theory.
Possibly, this could even help with supersymmetrizing the massive gravity potential.

\vspace{3pt}

\paragraph{Acknowledgements.} We are grateful to Ignatios Antoniadis , Ver\'onica Errasti-D\'iez,  
Fawad Hassan and Julio M\'endez-Zavaleta for useful discussions. 
We would like to especially thank James Bonifacio for his valuable comments on the draft. 
This work is supported by a grant from the Max-Planck-Society.

\appendix
\section{Independence of $\gmn$ and $\bmn$}\label{app:independence}

In this appendix we discuss the mild assumptions under which the vierbein combinations 
$\gmn$ and $\bmn$ can be treated as independent. 
Consider a gauge-fixed vierbein $E^a_{~\mu}$. The general vierbein is given by 
${e}^a_{~\mu}=\Lambda^a_{~b}E^b_{~\mu}$ where $\Lambda^a_{~b}$ is a local Lorentz tranformation
satisfying $\Lambda^\mathrm{T}\eta \Lambda=\eta$.
As already discussed in Ref.~\cite{Hassan:2012wt}, 
we can parameterize it by its Caley transform~\cite{Caley},
\beqn
\Lambda^a_{~b}=\big[(\eta-A)^{-1}\big]^{ac}\big[\eta_{cb}+A_{cb}\big]\,,
\eeqn
where $A_{ab}=-A_{ba}$ contains the 6 Lorentz parameters. We then have that,
\beqn
\gmn=E^a_{~\mu}\eta_{ab}E^b_{~\nu}\,,
\qquad
\bmn={\Lambda}^a_{~b}E^b_{~\mu}\eta_{ac}\te^c_{\nu}-{\Lambda}^a_{~b}E^b_{~\nu}\eta_{ac}\te^c_{\mu}\,.
\eeqn
The 6 components in $A_{ab}$ do not show up in the Lorentz invariant metric $\gmn$, whereas $\bmn$
depends on all of them. A field redefinition $\bmn\rightarrow A_{ab}$ would thus render the metric
and antisymmetric components independent. 
But it is not immediately obvious how to invert this dependence and express $A_{ab}$ as a function 
of $\bmn$ in a covariant way. Nevertheless, at the level of a 
linearized Lorentz transformation, we can easily derive the condition
for the invertibility of the dependence.

To this end, let us assume that the symmetric vierbein condition can be enforced by a 
local Lorentz tranformation $\bar{\Lambda}^a_{~b}$ of the general vierbein $e^a_{~\mu}$,\footnote{This
assumption is equivalent to the existence of the square-root matrix $\sqrt{g^{-1}f}$ with
$f=\te^\mathrm{T}\eta\te$~\cite{Deffayet:2012zc}, which appears in the tensor formulation of 
massive gravity~\cite{Hassan:2011vm}.}
\beqn\label{svcllt}
\bar{\Lambda}^a_{~b}e^b_{~\mu}\eta_{ac}\te^c_{\nu}
-\bar{\Lambda}^a_{~b}e^b_{~\nu}\eta_{ac}\te^c_{\mu}=0\,.
\eeqn
Infinitesimally we write the transformation as 
$\bar{\Lambda}_a^{~b}=\delta_a^{~b}-\bar{\omega}_{ac}\eta^{cb}$, such
that the above equation can be written in the form,
\beqn\label{bweq}
B_{\mu\nu}-e^a_{~\mu}\te^b_{~\nu}\bar{\omega}_{ab}+e^a_{~\nu}\te^b_{~\mu}\bar{\omega}_{ab}=0
\eeqn
Now we see that the right-hand side can be written as the invertible operator $P_{\mu\nu}^{~ab}$ 
(which we encountered in section~\ref{sec:locsym} when reproducing the local symmetry transformations)
acting on the infinitesimal Lorentz parameter $\bar{\omega}_{ab}$,
\beqn
P_{\mu\nu}^{~ab}\bar{\omega}_{ab}=2e^{[a}_{~[\mu}\te^{b]}_{~\nu]}\bar{\omega}_{ab}
=e^a_{~\mu}\te^b_{~\nu}\bar{\omega}_{ab}-e^a_{~\nu}\te^b_{~\mu}\bar{\omega}_{ab}\,.
\eeqn
Comparing to \eqref{bweq} we see that, 
\beqn\label{bomega}
P_{\mu\nu}^{~ab}\bar{\omega}_{ab}=\bmn\,.
\eeqn
In order to obtain the Lorentz transformation which satisfies \eqref{svcllt}, we need to invert this equation and
determine $\bar{\omega}_{ab}$ in terms of the components of $\bmn$.
In other words, the invertibility of $P$ makes it possible to map the 6 components of $\bmn$
onto the 6 Lorentz parameters $\bar{\omega}_{ab}$ which do not show up in $\gmn$. This implies
that $\bmn$ and $\gmn$ can be treated as independent objects precisely when the symmetric
vierbein condition can be satisfied by a Lorentz transformed $e^a_{~\mu}$.

Analogously, the operator $\tilde{P}_{\mu\nu}^{~ab}$ (which we encountered
in section~\ref{sec:eom} when separating the tensor equations) maps the Lorentz 
parameters which symmetrize the inverse $e_a^{~\mu}\eta^{ab}\te_b^{~\nu}$ onto the components of 
its antisymmetric part $\tilde{B}^{\mu\nu}=e_a^{~[\mu}\eta^{ab}\te_b^{~\nu]}$. Since the symmetry of
$e^a_{~\mu}\eta_{ab}\te^b_{~\nu}$ is equivalent to the symmetry of its inverse 
$e_a^{~\mu}\eta^{ab}\te_b^{~\nu}$, the local Lorentz transformation 
parametrized by $\bar{\omega}_{ab}$ is the same in both cases.
Thus we have that $\tilde{P}_{ab}^{~\mu\nu}\eta^{ac}\eta^{bd}\bar{\omega}_{cd}=\tilde{B}^{\mu\nu}$.

\section{Details of the ADM analysis}\label{app:adm}

This appendix contains technical details on the 3+1 split and 
degree of freedom counting performed in section~\ref{sec:ADM}.

\subsection{Standard massive theory for $\bmn$}\label{admmassb}

We begin by performing an ADM analysis of the standard massive theory for $\bmn$ in curved space,
whose Lagrangian reads,
\begin{equation}\label{massiveB}
\mathcal{L}_{B} = -  \frac1{2\cdot3!} \sqrt{g}~H_{\mu\nu\rho}H^{\mu\nu\rho}
-m^2\sqrt{g}~ B_{\mu\nu}B_{\rho\sigma}g^{\mu\rho}g^{\nu\sigma}\,.
\end{equation}
We introduce ADM variables to parametrize the metric and its inverse by~\cite{Arnowitt:1962hi},
\begin{equation}\label{ADMG}
g_{\mu\nu} = \begin{pmatrix}
-N^2 + N_k N^k & N_j \\ N_i & \gamma_{ij}
\end{pmatrix}\,, \qquad
g^{\mu\nu} = \frac1{N^2}\begin{pmatrix}
-1 & N^j \\ N^i & N^2\gamma^{ij} - N^iN^j
\end{pmatrix}
\end{equation}
where $N^i=\gamma^{ij}N_j$ and for the determinants we now have $\sqrt{g}=N\sqrt{\gamma}$. 
Moreover, we split the antisymmetric tensor $\bmn$ into its spatial components $B_{ij}$ 
and its mixed space-time components $B_{0j}$.
The canonical momentum conjugate to $B_{mn}$ is then obtained as,
\beqn\label{bcanm}
\Pi^{mn}\equiv\frac{\partial\mathcal{L}_B}{\partial \dot{B}_{mn}}
=\frac{\sqrt{\gamma}}{2N}~\Big(
\partial_0B_{ij} -3N^k\partial_{[k}B_{ij]} - 2\partial_iB_{0j}  \Big)\gamma^{i[m}\gamma^{n]j}\,,
\eeqn
while the components $B_{0j}$ remain nondynamical, i.e.~$\Pi^{0j}=0$.
With some manipulations we can now rewrite the Lagrangian in the form, 
\beqn
\mathcal{L}_B
&=&
\Pi^{mn}\dot{B}_{mn}  - \frac{N}{\sqrt{\gamma}}\Pi^{mn}\Pi_{mn} 
- 3N^{[k}\Pi^{ij]}\partial_kB_{ij}  - 2\Pi^{ij}\partial_iB_{0j} 
 -\frac14 N\sqrt{\gamma} \partial_kB_{ij}\partial_lB_{pq} \Xi^{kl,[ij],[pq]} \nn\\
&~& + m^2\frac{\sqrt{\gamma}}{N} \big[ 2 B_{0i}B_{0j} \gamma^{ij} 
- 4B_{0j}B_{ik}N^i\gamma^{jk} - N^2 B_{ij}B_{kl}g^{ik}g^{jl} \big]\,,
\eeqn
where we have defined the following spatial tensor combination,
\beqn\label{xidef}
\Xi^{kl,ij,pq} \equiv \gamma^{lk}\gamma^{pi}\gamma^{qj} 
+ \gamma^{lj}\gamma^{pk}\gamma^{qi} + \gamma^{li}\gamma^{pj}\gamma^{qk}\,.
\eeqn
Due to $\Pi^{0i}=0$, the $B_{0i}$ equations are constraints which we need to solve explicitly
in order to see the linearity of the final action 
in the lapse $N$ and the shift $N^i$. The solution is easily obtained and reads,
\begin{equation}
B_{0i} = N^jB_{ji} - \frac{N}{2m^2\sqrt{\gamma}}\partial_j\Pi^{jk}\gamma_{ki}\,.
\label{B0i}
\end{equation}
The final Lagrangian after integrating out the $B_{0j}$ then is, 
\beqn\label{massblag}
\mathcal{L}_B&=&
\Pi^{mn}\dot{B}_{mn}  
-N^i\Big(3 \Pi^{jk}\partial_{[i}B_{jk]} + 2B_{ij}\partial_k\Pi^{jk} \Big)
- N\left\{ \frac1{\sqrt{\gamma}}\Big( \Pi^{mn}\Pi_{mn} 
+ \frac1{2m^2}\partial_i\Pi^{ik}\partial_j\Pi^{jl}\gamma_{kl}\Big) \right. 
\nn\\
 &~& + \sqrt{\gamma}\Big( \tfrac14 \partial_kB_{ij}\partial_lB_{pq} 
 [\gamma^{lk}\gamma^{pi}\gamma^{qj} + \gamma^{lj}\gamma^{pk}\gamma^{qi} + \gamma^{li}\gamma^{pj}\gamma^{qk}]
 + m^2B_{ij}B_{kl}\gamma^{ik}\gamma^{jl}\Big)\left. \vphantom{ \frac1{\sqrt{\gamma}}} \right\}\,.
\eeqn
This is manifestly linear in $N$ and $N^i$, as expected since the Lagrangian in \eqref{massiveB} is 
diffeomorphism invariant.

Let us count the number of propagating degrees of freedom.
The components $B_{0j}$ have been algebraically determined by their own equations.
The dynamical objects are the spatial components $B_{mn}$, on which there are no further constraints.
Thus the massive theory for $\bmn$  defined by \eqref{massiveB} has 3 propagating degrees of 
freedom.\footnote{This result is of course in agreement with the covariant counting, which is particularly
easy to see in flat space. When $\gmn=\eta_{\mu\nu}$, the divergence
of the equations of motion for the massive $\bmn$ field gives the constraints $\partial^\mu\bmn=0$. 
Note that these are only 3 independent equations because 
$\partial^\nu\partial^\mu\bmn=0$ is identically satisfied.
The constraints therefore remove 3 out of the 6 components in $\bmn$, leaving 3 propagating modes.}
Adding the Einstein-Hilbert term to the action will introduce the 2 additional 
degrees of freedom of the massless graviton.

\subsection{Action with new mass term}
We now focus on our new action defined by \eqref{gbactionmass} and
show that it propagates the degrees of freedom of a massive $\bmn$
and a massive $\gmn$ field.
We parametrize the general vierbein $e^a_{~\mu}$ as
in equation \eqref{vbADM}
and we allow for a general form of $\te^a_{~\mu}$, which we write in $3+1$ form,
\beqn\label{ADMte}
\te^a_{~\mu} =
\begin{pmatrix}
L & 0\\
\tE^\alpha_{~j}L^j & \tE^\alpha_{~i}
\end{pmatrix} \, .
\eeqn
There is no need to include Lorentz parameters for the second vierbein.
They can all been shifted to the Lorentz matrix in $e^a_{~\mu}$, using the fact that a simultaneous
Lorentz transformation of both vierbeins leaves the action invariant. 

\subsubsection{Einstein-Hilbert term}

As is well-known, the Einstein-Hilbert term in the Lagrangian written in terms of 
the vierbein is of the form~\cite{Deser:1976ay} (see also the review~\cite{Peldan:1993hi}),
\begin{subequations}\label{EHadm}
\beqn
\mathcal{L}_e &=& \bar{\Pi}_\alpha^{~i}\dot{\bar{E}}^\alpha_{~i} - N\mathcal{C}^{(e)} - N^i\mathcal{C}^{(e)}_i\,, 
\eeqn
\end{subequations}
where $\bar{\Pi}_\alpha^{~i}$ is the canonical momenta conjugate to the spatial vierbein components ${\bar{E}}^\alpha_{~i}$.
The kinetic term for $\gmn$ is linear in $N$, and $N^i$ since the constraint contributions $\mathcal{C}^{(e)}$
and $\mathcal{C}^{(e)}_i $ depend only on the remaining ADM variables. Their precise form is irrelevant for the
arguments used in this work.

\subsubsection{Mass term}
We summarize briefly the relevant results of Ref.~\cite{Hinterbichler:2012cn}.
The mass term and the cosmological constant term in equation \eqref{gbactionmass} can be combined
into the potential,
\beqn
V 
&=& \sum_{n=0}^3 b_n\, 
\epsilon_{a_1a_2a_3a_4}\te^{a_1}\wedge\dots\wedge \te^{a_n}\wedge e^{a_{n+1}}\wedge\dots\wedge e^{a_4} \nn\\
&=& \sum_{n=0}^3 b_n\, 
\epsilon_{a_1a_2a_3a_4}\epsilon^{\mu_1 \mu_2\mu_3\mu_4} \te^{a_1}_{~\mu_1}\dots 
\te^{a_n}_{~\mu_n} e^{a_{n+1}}_{~\mu_{n+1}}\dots e^{a_4}_{~\mu_4}\,.
\eeqn
Note that there are 4 terms since we have included the cosmological constant $\Lambda$ for 
$e^a_{~\mu}$ by defining $b_0\equiv 2\Lambda/(m^24!)$. 
The ADM parametrizations of the two vierbeins in \eqref{vbADM} and \eqref{ADMte} give,
\begin{subequations}
\begin{align}
e^0_{~0} &= \Gamma(N + v_\alpha E^\alpha_{~i}N^i) \,,
\qquad ~~\te^0_{~0} = L\,, \\
e^\alpha_{~0} &= \Gamma v^\alpha N + \mathcal{V}^\alpha_{~\beta}E^\beta_{~i}N^i\,, 	
\qquad \te^\alpha_{~0} = \tE^\alpha_{~i}L^i\,, 
\end{align}
\end{subequations}
while all other (purely spatial) components are independent of $N$, $N^i$, $L$ and $L^i$.
The antisymmetric wedge product (i.e.~the factor of $\epsilon^{\mu_1 \mu_2\mu_3\mu_4}$)
in the potential ensures its linearity in the components $e^a_{0}$ and 
$\te^a_{~0}$ and thus the linearity in $N$, $N^i$, $L$ and $L^i$.
We can therefore write the potential in the form,
\beqn\label{massadm}
V=N\mathcal{C}^{(V)}+N^i\mathcal{C}^{(V)}_i+L\tilde{\mathcal{C}}^{(V)}+L^i\tilde{\mathcal{C}}^{(V)}_i
\,,
\eeqn
where $\mathcal{C}^{(V)}$, $\mathcal{C}^{(V)}_i$, $\tilde{\mathcal{C}}^{(V)}$ and 
$\tilde{\mathcal{C}}^{(V)}_i$ are all functions of the remaining ADM variables alone.

\subsubsection{Kinetic term for $\bmn$}
So far we have seen that the Einstein-Hilbert term and the mass potential are linear in the lapse
and the shift functions. We now show that this is also the case for the kinetic term of 
the antisymmetric components $\bmn=e^a_{~\mu}\eta_{ab}\te^b_{~\nu}-\te^a_{~\mu}\eta_{ab}e^b_{~\nu}$.

To this end, we again split the $\bmn$ into its spatial components $B_{ij}$ 
and its mixed space-time components $B_{0j}$.
The $3+1$ form of the kinetic term of $\bmn$ in \eqref{gbaction} 
is the same as in \eqref{massblag}. It reads,
\beqn\label{kinBADM}
\mathcal{L}_B
=
\Pi^{mn}\dot{B}_{mn}  - \frac{N}{\sqrt{\gamma}}\Pi^{mn}\Pi_{mn}  
&-& 3N^{[k}\Pi^{ij]}\partial_kB_{ij}  + 2\partial_i\Pi^{ij}B_{0j}  \nn\\
&-&\frac14 N\sqrt{\gamma}~ \partial_kB_{ij}\partial_lB_{pq} \Xi^{kl,[ij],[pq]} \,,
\eeqn
where $\Xi^{kl,ij,pq}$ was already defined in \eqref{xidef} and 
$\gamma_{ij}=\delta_{\alpha\beta}E^\alpha_{~i}E^\beta_{~j} = g_{ij}$ 
is the spatial part of our spacetime metric $g_{\mu\nu}$ as in \eqref{ADMG}
with determinant $\gamma\equiv\det\gamma_{ij}$.
The canonical momentum for the spatial components $B_{mn}$ is given by \eqref{bcanm}.
We remind the reader that the standard mass term for the $\bmn$ field is not contained in the action.

The definition of $\bmn$ in terms of the vierbeins gives the following ADM expressions,
\begin{subequations}
\beqn
B_{0i } &=& 
=   -\Gamma v_\alpha(L E^\alpha_{~i}  + N \tE^\alpha_{~i}) 
+  \tE^\alpha_{~k} \mathcal{V}_{\alpha\beta} E^\beta_{~l} (L^k\delta^l_i - \delta^k_iN^l) \,,\\
B_{ij} &=& 
= \mathcal{V}_{\alpha\beta} \tE^\alpha_{~[i} E^\beta_{~j]}
= \mathcal{V}_{\alpha\beta} R^{\beta}_{~\gamma}\tE^\alpha_{~[i}\bar{E}^\gamma_{~j]} \,.
\eeqn
\end{subequations}
Note that $B_{0i}$ is linear in the lapses and shifts while $B_{ij}$ is independent of them.
Inserting these expressions into \eqref{kinBADM}, we obtain,
\beqn
\mathcal{L}_B
=
\Pi^{mn}\dot{B}_{mn}  &-& \frac{N}{\sqrt{\gamma}}\Pi^{mn}\Pi_{mn}  
- 3N^{[k}\Pi^{ij]}\partial_k(\tE^\alpha_{~i}\mathcal{V}_{\alpha\beta} E^\beta_{~j}) \nn\\
&+& \partial_i\Pi^{ij}  \big[  -\Gamma v_\alpha(L E^\alpha_{~j}  + N \tE^\alpha_{~j})
+  \tE^\alpha_{~k} \mathcal{V}_{\alpha\beta} E^\beta_{~l} (L^k\delta^l_j - \delta^k_jN^l)  \big] \nn\\
&-&\frac14 N\sqrt{\gamma} \partial_k(\tE^\alpha_{~i}\mathcal{V}_{\alpha\beta} 
 E^\beta_{~j})\partial_l(\tE^\alpha_{~p}\mathcal{V}_{\alpha\beta} E^\beta_{~q}) \Xi^{kl,[ij],[pq]} \,.
\eeqn
This can be written in the form,
\beqn\label{bkinadm}
\mathcal{L}_B
&=&
\Pi^{mn}\dot{B}_{mn}  
-N\mathcal{C}^{(B)}-L\tilde{\mathcal{C}}^{(B)}-N^i\mathcal{C}^{(B)}_i-L^i\tilde{\mathcal{C}}^{(B)}_i \,,
\eeqn
where again none of $\mathcal{C}^{(B)}$, $\mathcal{C}^{(B)}_i$, $\tilde{\mathcal{C}}^{(B)}$
and $\tilde{\mathcal{C}}^{(B)}_i$ depend on $N$, $N^i$, $L$ and $L^i$.

\subsubsection{Full action in ADM variables}

Putting together the results for the two kinetic terms and the mass potential
obtained in the previous subsections, the whole action can be written in the form
\beqn\label{fullactadm}
S=\int\dd^4x\Big(\bar{\Pi}_\alpha^{~j}\dot{\bar{E}}^\alpha_{~j}+\Pi^{ij}\dot{B}_{ij} 
-N\mathcal{C} -N^i\mathcal{C}_i
-L\tilde{\mathcal{C}} -L^i\tilde{\mathcal{C}}_i\Big)\,,
\eeqn
where the $\mathcal{C}$ and $\mathcal{C}_i$, $\tilde{\mathcal{C}}$ and $\tilde{\mathcal{C}}_i$
contain the contributions from \eqref{EHadm}, \eqref{massadm} and \eqref{bkinadm}
which do not depend on any of the shifts or lapses.
Note that since $\te^a_{~\mu}$ is not a dynamical field in our setup, only the terms proportional
to $N$ and $N^i$ correspond to the constraints. The remaining terms make up
$\mathcal{L}_\mathrm{rest}$ in \eqref{almfinadmact}.


\end{document}